\documentclass[nofootinbib,twocolumn,nodate,showpacs,preprintnumbers,amsmath,amssymb,aps,prl]{revtex4-1}
\usepackage{amsmath}
\usepackage{latexsym}
\usepackage{amssymb}
\usepackage{color}
\newcommand*{\eps}{{\rlap{\lower2ex\hbox{$\,\,\tilde{}$}}{\epsilon_{ijk}}}}
\newcommand*{\EPS}{{\rlap{\lower2ex\hbox{$\,\,\tilde{}$}}{\epsilon_{i'j'k'}}}}
\newcommand*{\lmq}{{\rlap{\lower2ex\hbox{$\,\,\tilde{}$}}{\epsilon_{lmq}}}}
\newcommand*{\jmq}{{\rlap{\lower2ex\hbox{$\,\,\tilde{}$}}{\epsilon_{jmq}}}}
\newcommand*{\jql}{{\rlap{\lower2ex\hbox{$\,\,\tilde{}$}}{\epsilon_{jql}}}}
\newcommand*{\jlm}{{\rlap{\lower2ex\hbox{$\,\,\tilde{}$}}{\epsilon_{jlm}}}}
\newcommand*{\imq}{{\rlap{\lower2ex\hbox{$\,\,\tilde{}$}}{\epsilon_{imq}}}}
\newcommand*{\iql}{{\rlap{\lower2ex\hbox{$\,\,\tilde{}$}}{\epsilon_{iql}}}}
\newcommand*{\ilm}{{\rlap{\lower2ex\hbox{$\,\,\tilde{}$}}{\epsilon_{ilm}}}}
\newcommand*{\lmn}{{\rlap{\lower2ex\hbox{$\,\,\tilde{}$}}{\epsilon_{lmn}}}}
\newcommand*{\abc}{{\rlap{\lower2ex\hbox{$\,\,\tilde{}$}}{\epsilon_{abc}}}}
\newcommand*{\N}{{\rlap{\lower2ex\hbox{$\,\,\tilde{}$}}{N}}}
\newcommand{\tN}{{\rlap{\lower2ex\hbox{$\,\,\tilde{}$}}{N}}}
\newcommand*{\tM}{{\rlap{\lower2ex\hbox{$\,\,\tilde{}$}}{M}}}
\newcommand*{\imn}{{\rlap{\lower2ex\hbox{$\,\,\tilde{}$}}{\epsilon_{imn}}}}

\begin{document}
\title{Cosmic time and reduced phase space of General Relativity}

\author{Eyo Eyo Ita III}\email{ita@usna.edu}
\address{Physics Department, United States Naval Academy, 572C Holloway Road, Annapolis, Maryland 21402, USA}
\author{Chopin Soo}\email{Corresponding author: cpsoo@mail.ncku.edu.tw}
\address{Department of Physics, National Cheng Kung University, University Road, Tainan City 701, Taiwan}
\author{Hoi-Lai Yu}\email{hlyu@phys.sinica.edu.tw}
\address{Institute of Physics, Academia Sinica, Taipei 11529, Taiwan}
\input amssym.def
\input amssym.tex

\bigskip\bigskip

\begin{abstract}
\par\indent
 In an ever-expanding spatially closed universe, the fractional change of the volume is the preeminent intrinsic time interval to describe evolution in General Relativity. The expansion of the universe serves as a subsidiary condition which transforms Einstein's theory from a first class to a second class constrained system when the physical degrees of freedom (d.o.f.) are identified with transverse traceless excitations.
 The super-Hamiltonian constraint is solved by eliminating the trace of the momentum in terms of the other variables, and spatial diffeomorphism symmetry is tackled explicitly by imposing transversality. The theorems of Maskawa-Nishijima appositely relate the reduced phase space to the physical variables in canonical functional integral and Dirac's criterion for second class constraints to nonvanishing Faddeev-Popov determinants in the phase space measures. A reduced physical  Hamiltonian for intrinsic time evolution of the two physical d.o.f. emerges. Freed from the first class Dirac algebra, deformation of the Hamiltonian constraint is permitted, and natural extension of the Hamiltonian while maintaining spatial diffeomorphism invariance leads to a theory with Cotton-York term as the ultraviolet completion of Einstein's theory.
\end{abstract}
\maketitle
\section{Introduction}

   In this work two complementary perspectives will be adopted to frame the physical content of Einstein's theory in an ever-expanding spatially closed universe: 1)casting General Relativity (GR) as a {\it theory with second class constraints}\cite{Dirac},
   and 2)studying the {\it physical} canonical phase space functional integral of GR. In the process, we shall isolate the true d.o.f., solve all constraints, resolve the problem of time, and construct the corresponding reduced physical Hamiltonian. Freed from the shackles of the Dirac algebra and first class constraints, it will also be revealed that four-covariance is not really needed to capture the physical content of Einstein's theory\cite{SOOYU,SOOYU1,SOOITAYU,CS}; and modifications of GR are in fact allowed within the same framework.  This provides a rigorous canonical foundation for the consistency of Horava gravity theories\cite{Horava} which are obtained by modifying GR through its physical reduced Hamiltonian.

It is well-known that the canonical action of Einstein's General Relativity, $S=\int \int_{\Sigma}\Bigl(\tilde{\pi}^{ij}\frac{\partial{q}_{ij}}{\partial{t}}-N^iH_i-NH\Bigr)d^3x\,dt $, follows from the Einstein-Hilbert action and Arnowitt-Deser-Misner decomposition of the metric, $ds^2=g_{\mu\nu}dx^{\mu}dx^{\nu}=-N^2dt^2+{q}_{ij}\bigl(dx^i+N^idt\bigr)\bigl(dx^j+N^jdt\bigr)$, wherein the lapse and shift functions, $N$ and $N^i$, characterize the deformation of the spatial points from one hypersurface to the next. In this work the spatial manifold $\Sigma$ is assumed to be compact and without boundary to highlight the ``problem of time" and
its resolution. The EOM for the spatial metric, $q_{ij}$, relates the gravitational conjugate momentum to the extrinsic curvature, $K_{ij}=\frac{1}{2N}\Bigl(\frac{\partial{q}_{ij}}{\partial{t}}-\nabla_iN_j-\nabla_jN_i\Bigr) $, via $\tilde{\pi}^{ij}=\frac{\sqrt{q}}{2\kappa}\bigl(K^{ij}-q^{ij}K\bigr)$. The super-Hamiltonian and spatial diffeomorphism (or super-momentum) constraints (respectively $H=0$, $H_i=0$) form the first class Dirac algebra,
\begin{eqnarray}
\label{ALGEBRA}
&&\{H_i[{M^i}], H_j[{M'^j]}\}= H_k[(\mathcal L_{\vec{M}}\vec{M'})^k],\nonumber \\
&&\{H_i[{M^i}],H[M']\}=H[\mathcal L_{\vec{M}}M'],\nonumber\\
&&\{H[M],H[M']\} = H_i[q^{ij}(M\partial_jM'-M'\partial_jM)];
\end{eqnarray}
\noindent
wherein smearing functions are introduced in the Poisson brackets (for instance, ${H}_i[M^i] :=\int_{\Sigma} M^i H_i d^3x$).  Conversely, it has been shown the Dirac algebra is the hallmark of space-time 4-covariance and embeddability
of hypersurface deformations, from which the {\it explicit forms} of $H_i$ and $H$ of Einstein's theory of geometrodynamics can be regained\cite{Hojman}.

In the quantum context, following Dirac's own prescription, first class constraints can consistently be imposed to annihilate quantum states; but closure of the quantum algebra and divergences in Einstein's theory also need to be resolved meaningfully.  While $H_i$ can rigorously be interpreted as generators of spatial diffeomorphisms, the role of $H$ which is quadratic in the momentum is not as transparent, and the constraints do {\it not} generate 4-diffeomorphisms {\it off-shell}. What is clear is that {\it these constraints are needed to capture the physical content of Einstein's theory}, at least at the classical level. In what follows we shall recover this physical content from the two complementary viewpoints of  casting Einstein's GR as a  theory with second class constraints, and from studying the canonical phase space functional integral of GR.

In geometrodynamics, the fundamental variables, $(q_{ij},\widetilde{\pi}^{ij})$,  decompose in a manner which singles out the canonical pair $(\ln q^{\frac{1}{3}}$, $\widetilde{\pi})$ which commutes with
the remaining unimodular $\overline{q}_{ij}=q^{-\frac{1}{3}}q_{ij}$, and traceless momentum variable $\overline{\pi}^{ij}=q^{\frac{1}{3}}\bigl(\tilde{\pi}^{ij}-\frac{1}{3}q^{ij}\tilde{\pi}\bigr)$.  In fact the symplectic potential can be expressed as $\int {\tilde\pi}^{ij}\delta q_{ij} d^3x = \int ({\bar\pi}^{ij}\delta{\bar q}_{ij}  + {\tilde\pi}\delta\ln q^{\frac{1}{3}})\, d^3x$. For closed manifolds, Hodge decomposition of the spatial diffeomorphism scalar   $\delta \ln q^{\frac{1}{3}}=\frac{q^{ij}}{3}\delta q_{ij} $  or zero-form yields
$\delta \ln q^{\frac{1}{3}} = \delta{T}+\nabla_i{\delta y}^i$, wherein the spatially-independent $\delta T$ is spatial diffeomorphism invariant, whereas $\nabla_i{\delta y}^i$ can be gauged away since the change of $\ln q^{\frac{1}{3}}$ under spatial diffeomorphism is the Lie derivative ${\mathcal L}_{{{\overrightarrow N}}}\ln q^{\frac{1}{3}}  = \frac{2}{3}\nabla_i{N^i}$. In fact it follows\footnote{The Hodge decomposition is
$\delta\ln q^{\frac{1}{3}} = \frac{2}{3{\sqrt q}}\delta{\sqrt q}= \delta{T}+\nabla_i\delta{y}^i$. On multiplying by ${\sqrt q}$ and integrating over closed $\Sigma$, the $\delta y^i$ term drops out, yielding
$\frac{2}{3}{\delta V}= ({\delta T})V$, or $\delta T =\frac{2}{3}{\delta\ln V}$.} that $\delta T=\frac{2}{3}\frac{\delta V}{V} =\frac{2}{3}\delta\ln V$, wherein $V$ is the spatial volume of our universe\cite{SOOITAYU,CS}.
In an ever-expanding spatially closed universe, this serves as the preeminent and concrete physical time interval to discuss dynamics and evolution.

Intrinsic Time Gravity (ITG), or geometrodynamics with $\delta T$ intrinsic time interval and its corresponding Hamiltonian, has been advocated in a series of works\cite{SOOYU,SOOYU1,SOOITAYU,CS,CSYU,ZPEIJMPD,GRwave} which we shall briefly recap.
The main issues discussed in Ref.\cite{SOOYU} were the paradigm shift from full space-time covariance to spatial diffeomorphism invariance, unitary time development with gauge-invariant temporal ordering, emergence of classical space-time from constructive interference, and {\it a priori} versus {\it a posteriori} value of the lapse function in GR. A more rigorous proof of the emergent lapse function was given together with a discussion of Dyson time-ordering in intrinsic time in Ref.\cite{CS}.  The effect on the Hamiltonian structure of choosing intrinsic time slicings and simplifications in classical initial data construction were investigated in Ref.\cite {SOOYU1}, and the resultant generalized Lichnerowicz-York equation was shown to retain nice existence and uniqueness properties regardless of the additional Cotton-York term. In Ref.\cite{SOOITAYU}, a main ingredient was the introduction of Klauder's momentric variables\cite{Klauder} with its underlying $SU(3)$ group structure at each spatial point and the expression of the kinetic operator as a Casimir invariant. Asymptotic intrinsic time behavior of the theory was analyzed, together with its ground state and primordial quantum fluctuations. It was also pointed out that Cotton-York potential dominates at early times when the universe was small. Reference \cite{CSYU} highlighted the novel commutation relations introduced in \cite{SOOITAYU}, but the basic variables discussed therein were the eight components of the unimodular part of the spatial dreibein (rather than the five-component unimodular spatial metric) and the eight $SU(3)$ Klauder momentric variables, thus restoring eight components to each set of variables. The quantum Hamiltonian of intrinsic time gravity was elucidated in Ref.\cite{ZPEIJMPD}; in particular, heat kernel regularization was employed to demonstrate that, in addition to a Cotton-York term, Einstein's Ricci scalar potential emerged naturally from the simple positive-definite self-adjoint Hamiltonian. Consistent with the requirement of spatial compactness in the intrinsic time formulation,  GR waves in compact $k= +1$ cosmological de Sitter spacetine were studied\cite{GRwave}, and possible non-four-covariant Cotton-York contributions were shown to be negligible for long wavelengths at the current size of the universe, but these contributions can be important at early epochs when universe was much smaller.

In this work, the physical reduced phase space of GR will be analyzed within the context of intrinsic time formulation. With the addition of auxiliary condition for an ever-expanding closed universe,
the total restrictions no longer form a first class system, but constitute a perfect set of second class constraints in Dirac's terminology\cite{Dirac} when the two physical d.o.f. are identified with transverse-traceless excitations.
The super-Hamiltonian constraint is solved by eliminating $\tilde\pi$ in terms of the other variables, and spatial diffeomorphism symmetry is tackled explicitly by imposing transversality. A true physical Hamiltonian for intrinsic time evolution emerges; the theorems of Maskawa-Nishijima\cite{Maskawa} appositely relate the reduced phase space to the physical variables in canonical functional integral and Dirac's criterion for second class constraints to nonvanishing Faddeev-Popov determinants\cite{Popov} in the phase space measures.
Within this context, the further step of generalizing the form of the reduced Hamiltonian beyond GR is in fact permitted.  Natural extensions of the Hamiltonian while maintaining spatial diffeomorphism invariance lead to a theory with higher spatial curvature terms to improve ultraviolet convergence; thus this work also provides a firm canonical foundation for Horava gravity theories.

An ever-expanding spatially closed universe leads to the natural supplementary condition, which is a spatially independent $\delta \ln q^{\frac{1}{3}} =\delta{T}$, or  equivalenty,
\begin{eqnarray}
\label{ACTION8}
\chi =\frac{1}{3}\ln\Big[\frac{q}{q_0}\Big] -({T}-{T}_0) =0.
\end{eqnarray}
The super-Hamiltonian constraint is expressible in an interesting manner as\cite{SOOYU}
\begin{eqnarray}\label{GeneralH}
{\cal  H} &:=&\beta^2{\tilde\pi}^2 - {\bar{H}}^2 = -\frac{\sqrt{q}}{2\kappa} H=0,\nonumber \\
\bar {H}^2 &:=& { \bar{G}_{ijkl}\bar{\pi}^{ij}\bar{\pi}^{kl} + {\cal V}(\bar {q}_{ij}, q) }\nonumber \\
&=& { \frac{1}{2}[\bar{q}_{ik}\bar{q}_{jl} +\bar{q}_{il}\bar{q}_{jk}]\bar{\pi}^{ij}\bar{\pi}^{kl} + {\cal V}({q}_{ij})};
\end{eqnarray}
wherein Einstein's GR corresponds to ${\cal V}({\bar q}_{ij}, q) =- \frac{q}{(2\kappa)^2}(R - 2\Lambda_{\it{eff}} $)  and $\beta^2 := \frac{1}{6}$.
In addition, ${\bar H}(\ln q^{1/3},{\bar q}_{ij}, {\bar\pi}^{ij} )$ commutes with $\ln q^{\frac{1}{3}}$; so (\ref{ACTION8}) and (\ref{GeneralH}) lead to a remarkable consequence,
\begin{eqnarray}
\label{ACTION9}
\{\chi,{\cal H}\}=2\beta^2\tilde{\pi}= \pm 2\beta\bar{H},
\end{eqnarray}
on the constraint surface. Via the EOM,
\begin{equation}
\tilde\pi \propto {\sqrt q}K =\frac{3{\sqrt q}}{2N}(\frac{\partial{\ln q}^{\frac{1}{3}}}{\partial t} -\frac{2}{3}\nabla_iN^i),
 \end{equation}
 the trace of the extrinsic curvature; moreover, since $\lim_{\delta t \rightarrow 0}\frac{\delta\ln q^{\frac{1}{3}}}{\delta t} = \lim_{\delta t \rightarrow 0}\frac{\delta T}{\delta t} + \frac{1}{\sqrt q}\partial_i\frac{{\sqrt q}\delta y^i}{\delta t}$,  the latter term can always be compensated by choosing $\nabla_iN^i =\frac{3}{2}\nabla_i\frac{\partial y^i}{\partial t}$. Thus modulo spatial diffeomorphisms, and from a physical point of view, $\{\chi,{\cal H}\}$ is nonvanishing for an ever-expanding closed universe. This
is another motivation for casting GR as a theory with second class constraints, which we shall take up in the next section. Later on we shall discuss how to turn the {\it off-shell} value of ${\bar H}$ into a positive-definite entity.

Some may argue an ever-expanding universe is the {\it consequence} of Einstein's GR given the distribution of matter and the effective value of the cosmological constant, rather than an input.
However, current observations do not preclude the counterargument this work explores: an ever-expanding closed universe is fundamental to the framework of our physical universe and the resolution of the problem of time in classical and quantum GR.

\section{Second class constraints and phase space functional integral of GR}

As explained in Ref.\cite{Popov}, to reduce the phase space of a system with $2n$ d.o.f. subject to m constraints to a well-defined $2(n-m)$-dimensional canonical phase space with $(n-m)$ physical d.o.f., it is necessary to introduce m additional conditions.
These m conditions are often looked upon as ``gauge-fixing", but this terminology is somewhat obfuscating as these conditions (the transversality of physical gauge potentials as an example\footnote{For instance, in Electrodynamics, an arbitrary  gauge potential has the decomposition $A_i = A^T_i + \nabla_i\alpha$. Transversality of $A^T_i$ leads to $\nabla^2\alpha = \nabla^iA_i$; consequently $A^T_i = A_i - \nabla_i \frac{1}{\nabla^2} \nabla^jA_j$ is {\it explicitly  gauge invariant} under $A_i \mapsto A_i + \nabla_i\eta \quad\forall \eta $. So while the condition $\nabla^i A^{\rm phys}_i=0$ is often called ``gauge-fixing", the solution $A^{\rm phys}_i = A^T_i$ is, as shown, really {\it invariant} under gauge transformations. The Coulomb ``gauge" is thus a condition to extract the {\it gauge invariant physical d.o.f.} from the original configuration space.}) may stem from restrictions to gauge-invariant parts of natural decompositions, as well as from physical conditions compatible with empirical observations of the system. The crucial point is that the original constraints, supplemented by these extra restrictions can turn the system into one with $2m$ second class constraints and well-defined $(n-m)$ reduced canonical physical d.o.f.   For GR, an ever expanding spatially compact universe delivers the upshot of $\chi =0$, which, together with transversality of the unimodular perturbations, yield not only the correct 2 d.o.f., but also a nonvanishing physical Hamiltonian generating true (intrinsic) time translations.

To wit, we first note that the super-momentum constraint can be decomposed as  $H_i  = {\cal H}_i -\frac{2}{3}\nabla_i{\tilde\pi} $. Moreover,
$ -\frac{2}{3}\nabla_i{\tilde\pi}$ and  ${\cal H}_i$ separately, and respectively, generate spatial diffeomorphisms of $(\tilde\pi, \ln q^{\frac{1}{3}})$ and $({\bar\pi}^{ij}, {\bar q}_{ij})$; specifically,
$\int N^i{\cal H}_i d^3x =\int {\bar\pi}^{ij}\mathcal L_{\vec{N}}{\bar q}_{ij} d^3x$ with corresponding algebra,
\begin{eqnarray}
\{{\cal H}_i[{M^i}], {\cal H}_j[{M'^j]}\}&=& {\cal H}_k[(\mathcal L_{\vec{M}}\vec{M'})^k].
\end{eqnarray}
The explicit additional conditions $\chi_i =0$ which  will render $\{\chi, {\cal H}, , \chi_i, {\cal H}_i \}$ into a set of second class constraints will be addressed later; we first note the Poisson brackets of the set of constraints $\{\chi, {\cal H}, , \chi_i, {\cal H}_i \}$ are as tabulated below:
\begin{center}
\label{Pmatrix}
    \begin{tabular}{  l| l l lp{1.5cm}}
 $\{\,,\}$ &$ \,\,\chi $&\,${\cal H}$& $\,\,\,\,\chi_j $& $\,\,{\cal H}_l$ \\ \hline
   $\chi$&\,\,0& $\textstyle{2\beta^2\tilde{\pi}}$ &\,\,\, 0 &$\,\,\, 0$ \\
${\cal H}$&$\textstyle{-2\beta^2\tilde{\pi}}$  &\,\,\, X & $\,\,\,\,Y_j$& $\,\,Z_l$ \\
   $\chi_i$&\, 0 &$ -Y_i $&\,\,\, 0 &$\{\chi_i,{\cal H}_l\}$ \\
${\cal H}_k$& \,\,\,0 &  $-Z_k$  &  $\{{\cal H}_k, \chi_j\}$& $\{{\cal H}_k, {\cal H}_l\}$\\
    \end{tabular}
\end{center}
The matrix is invertible when $\{\chi, {\cal H}\}=2\beta^2\tilde{\pi}$ and $\det[\{\chi_i,{\cal H}_j\}] $ are both nonvanishing. Consequently, in Dirac's terminology, $\{\chi, {\cal H}, \chi_i, {\cal H}_i\}$ perfectly constitutes a set of ``second class constraints"\cite{Dirac}. Maskawa and Nakajima\cite{Maskawa} showed that any set of canonical variables governed by second class constraints is canonically equivalent to $ Q^r, P_r; {\cal Q}^\alpha , {\cal P}_\alpha$ such that the constraints read ${\cal Q}^\alpha ={\cal P}_\alpha  =0 $. Moreover, the Dirac bracket\cite{Dirac} is equal to the Poisson bracket calculated in terms of the reduced set of unconstrained variables i.e. $\{A, B\}_{\rm Dirac} = \frac{\partial A}{\partial Q^r}\frac{\partial B}{\partial P_r} - \frac{\partial B}{\partial Q^r}\frac{\partial A}{\partial P_r} $.  Thus all physics and computations become transparent and simplify greatly if the reduction to true d.o.f. can be completed and the reduced Hamiltonian constructed explicitly. In the canonical functional integral one can view  ${\cal Q}^\alpha = \chi, \chi_i$; and with ${\cal H}_\alpha = {\cal H}, {\cal H}_i$, it follows that
 $\det[\{{\cal Q}_{\alpha}, {\cal H}_\beta\}]  = \det[\frac{\partial{\cal H}_{\beta}}{\partial{\cal P}_\alpha}]\neq 0$ allows ${\cal H}_{\beta}=0$ to be solved in terms of ${\cal P}_\alpha$.
 Since
 \begin{equation}
 \delta({\cal Q}^{\alpha})\det[\{{\cal Q}^{\alpha},{\cal H}_\beta\}]\delta({\cal H}_{\beta}({\cal P}_{\alpha}))  = \delta({\cal Q}^{\alpha})\delta({\cal P}_{\alpha}),
 \end{equation}
 our previous factors, $\det[\{\chi, {\cal H}\}]$ and $\det[\{\chi_i, {\cal H}_j\}]$,  are just the required Faddeev-Popov determinants\cite{Popov} in $\det[\{{\cal Q}^{\alpha},{\cal H}_\beta\}]$ to enforce the restriction to the true physical phase space $(Q^r, P_r)$.

The canonical phase space functional integral is
\begin{eqnarray}
\label{ACTION10}
\hspace{-0.6cm}{\int}{D\mu}\Theta(-\tilde{\pi}) {\exp}\Bigl[{\tfrac{i}{\hbar}{\int}{dt}{\int}_{\Sigma}(\bar{\pi}^{ij}\tfrac{\partial\bar{q}_{ij}}
{\partial{t}}}+\tilde{\pi}\tfrac{\partial{\ln}q^{1/3}}{\partial{t}})d^3x\Bigr].
\end{eqnarray}
\noindent
In the functional integral we have inserted a theta function to pick out the choice of positive-definite $\bar H$  compatible with a expanding universe i.e.  $\tilde{\pi}=-\frac{\sqrt{q}K}{\kappa}=-\frac{\bar{H}}{\beta}$, whereas ${\cal H} =0$ is satisfied by $\tilde\pi =\pm\frac{\bar H}{\beta}$.
As the total determinant of the Poisson brackets in the table factorizes, the total measure will be associated with a direct product $\det[\{\chi, {\cal H}\}]\det[\{\chi_i,{\cal H}_j\}] $. To wit, including the determinants and constraints,  the complete measure is
\begin{eqnarray}
\label{ACTION11}
D\mu=D{\bar\mu}\prod_x{\delta\ln q^{\frac{1}{3}}\delta{\tilde \pi}}\,{\det}\{\chi ,{\cal H}\}\delta(\chi)\delta({\cal H}),
\end{eqnarray}
\noindent
with
 \begin{eqnarray}
 \label{ACTION111}
\hspace{-0.4cm}D{\bar\mu}=\prod_{x}{(2\pi\hbar)^3}\prod_{i,j}\tfrac{\delta\bar{\pi}^{ij}\delta\bar{q}_{ij}}{2\pi\hbar}\det[\{\chi_i, {\cal H}_j\}]\delta(\chi_i)\delta({\cal H}_j);
\end{eqnarray} consistent with the Faddeev-Popov prescription\cite{Popov}. In functional integrals, the condition $\chi_i=0$ and the Faddeev-Popov determinant can also be realized by integrating over auxiliary fields ${\tilde b}^i$  and complex Grassmannian ghost fields $({\bar c}^i , c^i)$  through $\int\, \prod_{k}\prod_{x\in\Sigma} \frac{d{\tilde b}^k }{2\pi}d{\bar c}^k dc^k  \exp({i\int\,{\tilde b}^i\chi_i + {\bar c}^i\{ \chi_i, {\cal H}_j\}c^j )d^3x'}$).

\subsection{A. Spatial diffeomorphisms}
In this subsection we address spatial diffeomorphisms and the resultant d.o.f. of the theory.  Supplementary to the ${\cal H}_i=0$ constraints and consistent with the gauge-invariance of transverse fluctuations, we introduce
$\chi_i= q^{\frac{1}{3}}{\nabla^*}^j{\delta{\bar q}}^{TT}_{ji}=0$ with ${\delta {\bar q}}^{TT}_{ij} = {\bar q}_{ij} -{\bar q}^{*}_{ij}$, wherein ${\nabla^*}$ denotes covariant derivative with respect to background ${q}^{*}_{ij}= q^{\frac{1}{3}}{\bar q}^*_{ij}$.  It follows from $ \bar{\pi}^{ij}:= q^{\frac{1}{3}}   [\tilde{\pi}^{ij}  - \frac{q^{ij}}{3}{\tilde\pi}] $ that the Poisson brackets for the barred variables are,

\begin{eqnarray}
\label{qT}
&&\{{\bar{q}}_{ij}(x),{\bar{q}}_{kl}(y)\} =0, \quad  \{{\bar{q}}_{kl}(x),{\bar{\pi}}^{ij}(y)\}= P^{ij}_{kl}\,\delta^3(x-y), \nonumber \\
&& \{{\bar{\pi}}^{ij}(x),{\bar{\pi}}^{kl}(y)\}=\frac{1}{3}({\bar q}^{kl}{\bar{\pi}}^{ij} -{\bar q}^{ij}{\bar{\pi}}^{kl})\delta^3(x-y);
\end{eqnarray}
with $P^{ij}_{kl} :=  \frac{1}{2}(\delta^i_k\delta^j_l + \delta^i_l\delta^j_k) - \frac{1}{3}{\bar{q}}^{ij}\bar{q}_{kl}$ denoting the traceless projection operator.

Under spatial diffeomorphisms, the unimodular metric variable changes  by
\begin{eqnarray}
\label{DIFFEO1}
{\mathcal L}_{\vec\xi}\,{\bar q}_{ij}&=&\xi^k\partial_k {\bar q}_{ij} + {\bar q}_{ik}\partial_j\xi^k +{\bar q}_{kj}\partial_i\xi^k -\frac{2}{3}{\bar q}_{ij}\partial_k\xi^k\nonumber \\ &=&q^{-\frac{1}{3}}(\nabla_i\xi_j+\nabla_j\xi_i-\frac{2}{3}q_{ij}\nabla^k\xi_k) \nonumber\\
&=&2q^{-\frac{1}{3}}P^{lk}_{ij}\nabla_l\xi_k = : {\cal L}_{(ij)}^{\quad {k}}\xi_k
\end{eqnarray}
wherein $\nabla$ is the Levi--Civita connection of the spatial metric.

We may introduce the positive-definite block-diagonal metric $G^{IJ} = \left[
                                                   \begin{array}{cc}
                                                     {\bar G}^{ijkl} & 0 \\
                                                     0 & q^{mn} \\
                                                   \end{array}
                                                 \right]$
wherein $I= ((ij), m)$, $J=((kl), n)$, and  ${\bar G}^{ijkl} =\frac{1}{2} ({\bar q}^{ik}{\bar q}^{jl} +{\bar q}^{il}{\bar q}^{jk})$.  This allows the definition a positive-definite inner product
\begin{eqnarray}
\label{PBgauge1}
&&\langle  U^1, U^2\rangle :=\int {\bar G}^{IJ}U_I U'_J {\sqrt q}d^3x \nonumber \\
&=& \int {\bar G}^{ijkl}\delta{\bar q}^1_{ij} \delta{\bar q}^2_{kl} {\sqrt q}d^3x + \int {q}^{mn}\xi^1_{m}\xi^2_{n} {\sqrt q}d^3x
\end{eqnarray}
for any pair $U^{1,2}_I = \left[
                                                              \begin{array}{cc}
                                                                \delta {\bar q}^{1,2}_{ij} \\
                                                                \xi^{1,2}_k \\
                                                              \end{array}
                                                            \right]$.
And associated with spatial diffeomorphisms  is the operation
${\cal L}_I\,^J\xi_J = \left[
                                                   \begin{array}{cc}
                                                     0 &  {\cal L}_{(ij)}^{\quad m}\\
                                                     1 & 0 \\
                                                   \end{array}
                                                 \right]\cdot\left[
                                                              \begin{array}{cc}
                                                                0 \\
                                                                \xi_m \\
                                                              \end{array}
                                                            \right] = \left[
                                                              \begin{array}{cc}
                                                                {\cal L}_{\vec\xi}{\bar q}_{ij} \\
                                                                0\\
                                                              \end{array}
                                                            \right]  $.
To demonstrate that the determinant of the relevant Poisson bracket is almost generically nonvanishing for an arbitrary background, we note that
\begin{widetext}
\begin{eqnarray}
\label{PBgauge}
\Big\{ \int \zeta^m \chi_m{\sqrt q}d^3x , \int \xi^n {\cal H}_n d^3y\Big\} &=&\Big\{\int \zeta^m{\sqrt q}q^{\frac{1}{3}}\nabla^{*n}({\bar q}_{mn} - {\bar q}^*_{mn}) d^3x , \int ({\cal L}_{\vec \xi}{\bar q}_{ij})^* {\bar \pi}^{ij}d^3y\Big\}\\ \nonumber
&=&-\frac{1}{2}\int {\sqrt q}q^{\frac{1}{3}}(\nabla^{*i}\zeta^j + \nabla^{*j}\zeta^i -\frac{2}{3}{q}^{ij}\nabla^*_k\zeta^k) ({\cal L}_{\vec \xi}{\bar q}_{ij})^* d^3y\\ \nonumber
&=&-\frac{1}{2}\int {\bar G}^{*ijkl}({\cal L}_{(ij)}^{\quad m}\zeta_m )^*({\cal L}_{(kl)}^{\quad n}\xi_n)^* {\sqrt q}d^3y\\ \nonumber
&=&-\frac{1}{2}\int {\bar G}^{*IJ}({\cal L}_{I}^{*K}\zeta_K )({\cal L}_{J}^{*L}\xi_L) {\sqrt q}d^3y =-\frac{1}{2}\langle {\cal L}^{*}\zeta, {\cal L}^* \xi\rangle\\ \nonumber
&=&-\frac{1}{2} \langle \zeta, {\cal L}^{*\dagger}{\cal L}^* \xi\rangle;
\end{eqnarray}
\end{widetext}
which means that the determinant of $\{ \chi_m, {\cal H}_n \}$ is that of a negative semidefinite operator. It follows that a zero mode is present iff ${\cal L}_{(ij)}^{*\quad {k}}\xi_k =0$, equivalently
 $\xi^i$ must be a conformal Killing vector which obeys ${\cal L}_{\vec\xi} q^*_{ij}=\frac{2}{3}q^*_{ij}\nabla^*_k\xi^k$.
 Changes due to spatial diffeomorphisms are orthogonal to transverse-traceless physical excitations since
\begin{eqnarray}
\label{orthogonal}
&&\int {\bar G}^{ijkl}\delta{\bar q}^{TT}_{kl}({\cal L}_{\vec\xi}{\bar q}_{ij}){\sqrt q}d^3x \nonumber \\
&=&\int {\bar G}^{ijkl}\delta{\bar q}^{TT}_{kl} q^{-\frac{1}{3}}(\nabla_i\xi_j+\nabla_j\xi_i-\frac{2}{3}q_{ij}\nabla^m\xi_m){\sqrt q} d^3x\nonumber \\
&=&0
\end{eqnarray}
after integrating by parts and invoking $\nabla^k\delta{\bar q}^{TT}_{kl} =0$. The orthogonal decomposition into physical and gauge changes  $\delta{\bar q}_{ij} =\delta{\bar q}^{TT}_{ij} +{\cal L}_{\vec\xi}{\bar q}_{ij}$ does not fix
${\vec\xi}$ uniquely if one or more zero modes, ${\vec\xi}_{o\alpha}$, exist; but even then the physical $\delta{\bar q}^{TT}_{ij}$ is unaffected since these conformal Killing vectors  satisfy ${\cal L}_{{\vec\xi}_{o\alpha}}{\bar q}_{ij} =0$. This parallels the arguments given by York\cite{York2} for the decomposition of the momentum variable which we shall briefly recap. To wit, any arbitrary momentum can be expressed as
\begin{eqnarray}
\label{decom}
{\tilde\pi}^{ij}&=&{\tilde\pi}^{ij}_{TT} + \frac{q^{ij}}{3}{\tilde\pi} +\sqrt{q} (LW)^{ij},\nonumber\\
\qquad (LW)^{ij} &:=& \nabla^{i} W^{j}+\nabla^{j} W^{i} -\frac{2q^{ij}}{3}\nabla_kW^k;
\end{eqnarray}
and the requirement
   \begin{equation}
   \label{DiffeoW}
  (\triangle_L W)^j := \nabla_i(L W)^{ij}  = \frac{1}{\sqrt q}\nabla_i({\tilde\pi}^{ij} - \frac{q^{ij}}{3}{\tilde\pi})
   \end{equation}
follows from $\nabla_i {\tilde\pi}^{ij}_{TT}=0$. Thus $W^i$ depends only on the traceless part of ${\tilde\pi}^{ij}$; moreover, $(LW)^{ij}$ is also traceless.
The operator $\triangle_L$ is strongly elliptic and its kernel consists of those $W^i_{o\alpha}$ which satisfy $(LW_{o\alpha})^{ij} =0$. The general solution, $W^i = W^i_p + \sum_{\alpha}c^\alpha W^i_{o\alpha}$,
is a linear combination of the particular solution and elements of the kernel. But $(LW)^{ij}  =  (LW_p)^{ij}$, so the presence of a nontrivial kernel does not affect the physical mode ${\bar\pi}^{ij}_{TT}$ and the uniqueness of the decomposition (\ref{decom}). By the same arguments, the presence of zero modes in (\ref{PBgauge}) does not disturb  $\delta{\bar q}^{TT}_{ij}$. Note that the full diffeomorphism constraint
 $H_i :=-2q_{ik}\nabla_j{\tilde\pi}^{jk}_{\rm phys} =0$ is actually satisfied by ${\tilde\pi}^{ij}_{\rm phys} = {\tilde\pi}^{ij}_{TT}$.

 Substituting the decomposition of ${\tilde\pi}^{ij}$ into the symplectic potential and integrating by parts terms with $W^i$ reveal that
  \begin{eqnarray}\label{DOF}
 {\textstyle\int}{\tilde\pi}^{ij}\delta q_{ij}d^3x &=&{\textstyle\int}({\bar\pi}^{ij}_{TT}\delta{\bar q}_{ij}- 2W^jq^{\frac{1}{3}}\nabla^i\delta{\bar q}_{ij}+\pi\delta\ln q^{\frac{1}{3}})d^3x;  \nonumber \\
 &&\qquad {\bar\pi}^{ij}_{TT}:={\pi}^{ij}_{TT}q^{\frac{1}{3}}.
\end{eqnarray}
When restricted to the physical subspace with $\nabla^i\delta{\bar q}^{\rm phys.}_{ij} =0$ i.e. to traceless-transverse excitations $\delta{\bar q}^{{TT}}_{ij}$, the
 symplectic potential reduces to $ \int \,({\bar\pi}^{ij}_{TT}\delta{\bar q}^{{TT}}_{ij}+ \pi \delta\ln q^{\frac{1}{3}})$ yielding two physical TT degrees of freedom,
 and an extra pair $(\ln q^{\frac{1}{3}}, \pi)$ to feature in intrinsic time and Hamiltonian density.
For perturbations about any background $q^*_{ij}= q_{ij}-\delta q_{ij}$, the linearized physical spatial metric modes $\delta{\bar  q}^{\rm phys}_{ij} =
(P^{kl}_{ij})^*\delta q_{kl} $ are traceless ($q^{*ij}\delta{\bar q}^{\rm phys}_{ij} =0$) and transverse
(${\nabla}^ {*i}\delta {\bar q}^{\rm phys}_{ij} =0)$ w.r.t $q^*_{ij}$, correctly accounting for the perturbative graviton degrees of freedom.

In Electrodynamics, the physical d.o.f. is the transverse projection $A^T_i = (\delta^j_i - \nabla_i \frac{1}{\nabla^2} \nabla^j)A_j$ which is {\it gauge invariant} under $A_i \mapsto A_i + \nabla_i\eta \quad\forall \eta $.
The GR analogy can be made concrete: Eq.(\ref{orthogonal}) implies ${\cal L}^\dagger\cdot\left[
                                                                               \begin{array}{c}
                                                                                 \delta{\bar q}^{TT}_{ij} \\
                                                                                 0\\
                                                                               \end{array}
                                                                             \right]$ vanishes. By decomposing  $\delta{\bar q}_{ij} =\delta{\bar q}^{TT}_{ij} +{\cal L}_{\vec\eta}{\bar q}_{ij}$, or equivalently, $\left[
                                                                               \begin{array}{c}
                                                                                 \delta{\bar q}_{ij} \\
                                                                                 0\\
                                                                               \end{array}
                                                                             \right] =\left[
                                                                               \begin{array}{c}
                                                                                 \delta{\bar q}^{TT}_{ij} \\
                                                                                 0\\
                                                                               \end{array}
                                                                             \right] + {\cal L}\cdot\left[
                                                                               \begin{array}{c}
                                                                                 0 \\
                                                                                 \eta_k\\
                                                                               \end{array}
                                                                             \right]$, and acting with ${\cal L}^\dagger $ on the equation,
                                                                             it follows that $\left[\begin{array}{c}
                                                                                 0 \\
                                                                                 \eta_k\\
                                                                               \end{array} \right]= ({\cal L}^\dagger{\cal L})^{-1}\cdot {\cal L}^\dagger\cdot\left[
                                                                               \begin{array}{c}
                                                                                 \delta{\bar q}_{ij}\\
                                                                                 0\\
                                                                               \end{array}\right]$. Substituting this back into the decomposition yields the physical $\left[\begin{array}{c}
                                                                                 \delta{\bar q}^{TT}_{ij} \\
                                                                                 0\\
                                                                               \end{array}
                                                                             \right] =\Big(I - {\cal L}\cdot ({\cal L}^\dagger{\cal L})^{-1}\cdot {\cal L}^\dagger\Big)\cdot \left[\begin{array}{c}
                                                                                 \delta{\bar q}_{ij} \\
                                                                                 0\\
                                                                               \end{array}
                                                                             \right]$, which is {\it explicitly invariant} under spatial diffeomorphisms  $\left[\begin{array}{c}
                                                                                 \delta{\bar q}_{ij} \\
                                                                                 0\\
                                                                               \end{array}
                                                                             \right]\mapsto \left[\begin{array}{c}
                                                                                 \delta{\bar q}_{ij} \\
                                                                                 0\\
                                                                               \end{array}
                                                                             \right] + {\cal L}\cdot\left[\begin{array}{c}
                                                                                 0\\
                                                                                 \xi_k\\
                                                                               \end{array}
                                                                             \right] \quad \forall\, \xi_k $. A similar decomposition and projection can be carried out for $\bar\pi^{ij}_{TT}$. These true d.o.f. are physical observables.

\subsection{B. Physical Hamiltonian, and generalization beyond Einstein's theory}
Reduction of the canonical functional integral to the {\it unconstrained} physical phase space can now be completed.
Taking (\ref{ACTION8}) and (\ref{ACTION9}) into account, (\ref{ACTION10}) leads, upon integrating over $D\mu$ of (\ref{ACTION11}), to
\begin{eqnarray}
\label{ACTION12}
\hspace{-0.5cm}\int D{\bar\mu}
\, {\exp}\Bigl[{\frac{i}{\hbar}\int\int_{\Sigma}(\bar{\pi}^{ij}\tfrac{\partial\bar{q}_{ij}}{\partial{T}}}-\tfrac{{\bar H}(q(T),{\bar q}_{ij}, {\bar \pi}^{ij})}{\beta})d^3x{dT}\Bigr],
\end{eqnarray}
\noindent
with the {\it emergence of a Hamiltonian} generating $T$-translations. Further integration over $D{\bar\mu}$ of (\ref{ACTION111}) yields the true physical functional integral, $\int D{\bar\mu}_{\rm phys}\exp({\frac{i}{\hbar}S_{\rm phys}})$, with
\begin{eqnarray}
\label{ACTION14}
S_{\rm phys}&=&\int ({\int_{\Sigma}\bar{\pi}^{ij}_{TT}\tfrac{\partial{\bar q}^{TT}_{ij}}{\partial{T}}}d^3x)d{T}-\int H_{\rm phys}dT,\\
D{\bar\mu}_{\rm phys}&=&\prod_{x\in\Sigma}\prod_{i,j}\frac{\delta\bar{\pi}^{ij}_{TT}(x)\delta\bar{q}^{TT}_{ij}(x)}{2\pi\hbar};
\end{eqnarray}
and corresponding emergent {\it physical} Hamiltonian,
\begin{eqnarray}
\label{ACTION15}
H_{\rm phys}=\frac{1}{\beta}\int_{\Sigma}{\bar{H}(\ln q^{1/3}(T), {\bar q}^{TT}_{ij}, {\bar\pi}^{ij}_{TT})\, d^3x}.
\end{eqnarray}

A crucial point to note is that the precise form of $X, Y_i, Z_i$ play no role in total determinant of the Poisson brackets of the constraints discussed earlier, and in the above derivation of the physical Hamiltonian;
so changes in the precise form of ${\cal V}$ are in fact allowed within this framework.  Einstein's General Relativity (with $\beta= \frac{1}{\sqrt{6}}$ and ${\mathcal V} =- \frac{q}{(2\kappa)^2}[R - 2\Lambda_{\rm{eff}} $]) {\it is thus a particular realization of this wider class of theories} which is {\it compatible with spatial diffeomorphism symmetry and an ever-expanding closed universe}\cite{SOOYU, SOOYU1,CS}.
Generalization of ${\cal H} =\beta^2{\tilde\pi}^2 - {\bar H}^2$  to other cases with
 \begin{equation}
\label{INTRINSIC1}
{\bar H} = {\sqrt{ {\bar \pi^{ij}}  {\bar G_{ijkl}} {\bar \pi^{kl}} + \mathcal{V}[q_{ij}]}},
\end{equation}
wherein ${\mathcal V}(q,\bar q_{ij})$  is a spatial scalar density of weight two is allowed\cite{SOOYU}. In particular, the advantages of $\bar H = \sqrt{ {\hat Q}^{\dagger i}_{j}{\hat Q}^{j}_{i}+ q{\cal K}}$ with positive coupling ${\cal K}$ have been elaborated elsewhere\cite{SOOYU,SOOITAYU,ZPEIJMPD}. This also completes the earlier nonvanishing requirement of $\{\chi,{\cal H}\}=2\beta^2\tilde{\pi}= - 2\beta\bar{H}$.

In the quantum context, this Hamiltonian density is self-adjoint. Explicit spatial diffeomorphism invariance is achieved by introducing the interactions through ${\hat Q}^{i}_{j} := e^{W_T}{{\hat{\bar \pi}}}^{i}_{j}e^{-W_T}$ with
  $W_T$ being  a combination of the Einstein-Hilbert action, $W_{EH}$, and the Chern-Simons functional, $W_{CS}$, in three spatial dimensions i.e.
  \begin{eqnarray}
  W_T&=&W_{CS} + W_{EH} \nonumber\\
  &=&\frac{g}{4}\int{\tilde\epsilon}^{ijk}({\Gamma}^l_{im} \partial_j{\Gamma}^m_{kl} +\frac{2}{3}{\Gamma}^l_{im}{\Gamma}^m_{jn}{\Gamma}^n_{kl})\,d^3x\nonumber\\
  &&- \alpha\int {\sqrt q}Rd^3x.
  \end{eqnarray}
 As demonstrated in Ref.\cite{ZPEIJMPD} the final Hamiltonian in the limit of regulator removal is
 \begin{widetext}
\begin{equation}
\label{Hphys}
{H}_{\rm phys}=\int\frac{{\bar H}(x)}{\beta}d^3x, \quad \bar H = \sqrt{ {\hat Q}^{\dagger i}_{j}{\hat Q}^{j}_{i}+ q{\cal K}}
=\sqrt{{\bar{\pi}}^{\dagger i}_j {\bar{\pi}}^{j}_i+\hbar^2g^2\tilde{C}^i_j\tilde{C}^j_i - \frac{q}{(2\kappa)^2}(R - 2\Lambda_{\rm eff})}.
\end{equation}
\end{widetext}
wherein $\tilde{C}^{ij} = \frac{\delta W_{CS}}{\delta q_{ij}}$ is the Cotton-York tensor (density). Associated with the dimensionless coupling constant $g^2$ in (\ref{Hphys}), the Cotton-York term (which contains up to six spatial derivatives) modifies the propagator of Einstein's theory and ensures ultraviolet convergence\cite{Horava}.

The Cotton-York extension in (\ref{Hphys}) is analogous to the Yang-Mills magnetic field contribution in the Hamiltonian density,
$\frac{q_{ij}}{2}( {\hat{\tilde\pi}^{ia}_A}{\hat{\tilde\pi}^{ja}_A}+ \hbar^2{\tilde B}^{ia} {\tilde B}^{ja} )= \frac{q_{ij}}{2}{{\hat Q}^\dagger}\,^{ia}{\hat Q}^{ja}$,
with ${\hat Q}^{ia} = e^{W_{CS}}{\hat{\tilde \pi}^{ia}_A}e^{-W_{CS}} = {\hat{\tilde\pi}^{ia}_A} +i\hbar{\tilde B}^{ia} $; wherein   ${\hat{\tilde\pi}^{ia}_A} =\frac{\hbar}{i}\frac{\delta}{\delta A_{ia}}$ is the conjugate momentum to the gauge potential, and  the magnetic field
 ${\tilde B}^{ia} = \frac{\delta W_{CS}}{\delta A_{ia}}$ is the functional derivative of the Chern-Simons  functional of the Yang-Mills connection $A_{ia}$. In the case of Yang-Mills, there is no analog of the  Einstein-Hilbert action term, $\int {\sqrt q}Rd^3x$,  which can be added to $W_T$.

\section{Further remarks on extrinsic, intrinsic, and scalar field time}

In the extrinsic time formulation\cite{extrinsic}, constancy of $p:=\frac{{\tilde\pi}}{\sqrt q}$ is invoked, and the Hamiltonian constraint $H=0 $ translates into the Lichnerowicz-York equation\cite{Niall}, $p^2 - \frac{q}{\beta^2}{\bar H}^2({\bar\pi}^{ij}, {\bar q}_{ij}, q) = 0$. This determines $q$ uniquely, thus eliminating the d.o.f. $(\ln q^{\frac{1}{3}}, {\tilde\pi})$ in terms of $({\bar\pi}^{ij}, {\bar q}_{ij}, p)$. However, the reduced Hamiltonian (from $\int\int {\tilde\pi}\frac{\partial\ln q^{\frac{1}{3}}}{\partial t} dtd^3x = -\frac{2}{3}\int (\int {\sqrt  q}d^3x)dp)$  is then proportional to the spatial volume, wherein $q({\bar\pi}^{ij}, {\bar q}_{ij}, p)$ is known only implicitly and non-locally from the Lichnerowicz-York equation. In contradistinction, Intrinsic Time Gravity utilizes $d\ln q^{\frac{1}{3}}= dT$ as the universal time interval compatible with an expanding spatially closed universe, together with Hamiltonian constraint, $\beta{\tilde\pi} = -{\bar H}$. The resultant Hamiltonian is both {\it explicit}, and of the remarkable {\it relativistic form}, $H_{\rm phys.} =\frac{1}{\beta}\int {\bar H}d^3x= \frac{1}{\beta}\int  {\sqrt{ {\bar G_{ijkl}}{\bar \pi}^{ij}{\bar \pi}^{kl} + \mathcal{V}}}\,d^3x $.  Einstein's theory corresponds to $\beta= \frac{1}{\sqrt{6}}$ and ${\mathcal V} =- \frac{q}{(2\kappa)^2}[R - 2\Lambda_{\rm{eff}}]$, but, as explained, the scheme permits additional terms in the potential to improve ultraviolet convergence\cite{Horava}, while infrared divergence is completely curbed by spatial compactness.

A scheme which uses a real scalar field as ``time" would correspond to $\chi =\phi - k =0$ and ${\cal H}=\beta^2{\tilde\pi}^2 - {\bar H}^2 -\frac{1}{2\kappa}[\frac{{\tilde\pi}^2_{\phi}}{2}+\frac{1}{2}qq^{ij}\nabla_i\phi\nabla_j\phi +V(\phi)] =0$, with resultant $\{\chi, {\cal H}\}= -\frac{1}{2\kappa}{\tilde\pi}_{\phi}\approx \mp\frac{1}{\sqrt{2\kappa}}\sqrt{{2}(\beta^2{\tilde\pi}^2 - {\bar G_{ijkl}}{\bar \pi}^{ij}{\bar \pi}^{kl} - \mathcal{V}[{q}_{ij}])-\frac{1}{2\kappa}V|_{\phi=k} }$ on the constraint surface.
Among other potential problems, the negative-(semi)definite entity $-{\bar G_{ijkl}}{\bar \pi^{ij}}  {\bar \pi^{kl}}$ compromises the reality of ${\tilde\pi}_{\phi}$, and this carries over to the reduced Hamiltonian term, $\int\int {\tilde\pi}_{\phi}\frac{d\phi}{dt}dt\,d^3x =\int (\int {\tilde\pi}_{\phi} d^3x) dk$,  in the action. Moreover, unlike both the extrinsic and intrinsic time formulations above, it fails to eliminate the $(\ln q^{\frac{1}{3}}, {\tilde\pi})$ d.o.f., so while the total number of d.o.f. modulo constraints and subsidiary conditions is preserved, there are nevertheless 3 (and not 2) remaining {\it gravitational} d.o.f.. Incorporating Yang-Mills and matter (both scalar and fermionic) fields into the theory changes the total Hamiltonian constraint to
\begin{equation}
\label{total}
{\cal H}=\beta^2 {\tilde\pi}^2-{\bar H}^2 -H_{\rm{matter+YM}} =0.
\end{equation} But no couplings to ${\tilde\pi}$ appear in the usual $H_{\rm{matter+YM}}$, so the Poisson bracket in (\ref{ACTION9}) with the new ${\cal H}$ is unaffected. It follows ${\bar H}$ is modified to $\sqrt{\bar{H}^2+H_{\rm{matter+YM}}}$ in the reduced physical Hamiltonian of Eq.(\ref{ACTION15}). The super-momentum will now include the generator of spatial diffeomorphisms ${\tilde D}_i $ for these additional fields  i.e. ${H}^{\rm GR + matter}_i = -2q_{ik}\nabla_j{\tilde\pi}^{jk} + {\tilde D}_i $. The decomposition (\ref{decom}) of the generic momentum, ${\tilde\pi}^{ij} = {\tilde\pi}^{ij}_{TT} +  \frac{q^{ij}}{3}{\tilde\pi}  +{\sqrt q}(LW)^{ij}$, is still valid. The solution of the diffeomorphism constraint ${H}^{\rm GR + matter}_i =0$ is then ${\tilde\pi}^{ij}_{\rm phys} ={\tilde\pi}^{ij}_{TT} +  \sqrt{q}(LW_{\rm phys})^{ij}$, with the particular solution $W^i_{\rm phys}$ completely determined by\footnote{The analog in Electrodynamics is $E^i_{\rm phys} = E^i_T -\nabla^i\phi_{\rm phys}$, with $\phi_{\rm phys}$ satisfying the Poisson equation $\nabla^2\phi_{\rm phys} = -4\pi\rho$; consequently, $\nabla_i E^i_{\rm phys} = 4\pi\rho$. In the absence of sources the physical electric field is purely transverse.} $(\triangle_L W_{\rm phys})^i= \frac{1}{2\sqrt q}{\tilde D}^i$.
 This changes the explicit particular solution but does not disturb  ${\tilde\pi}^{ij}_{TT}$ at all (in pure GR,  ${\tilde D}_i =0$, yielding $W^i_{\rm phys} =0$ and ${\tilde\pi}^{ij}_{\rm phys} ={\tilde\pi}^{ij}_{TT}$). Thus inclusion of matter and Yang-Mills content does not alter the salient fact the unconstrained gravitational initial data lie in the transverse-traceless part of the momentum. In this reduction scheme, ${\tilde\pi}^{ij}_{\rm phys}$ is traceless even when $W^i_{\rm phys}$ is nontrivial in the presence of non-gravitational fields, while the freedom in the trace is eliminated by solving the Hamiltonian constraint as $\tilde\pi =-\frac{1}{\beta}\sqrt{ {\bar H}^2 + H_{\rm{matter+YM}}}$.
 That $H_{\rm phys}$ is ultimately $T$-dependent may be disconcerting at first, but this feature also occurs in the York formulation wherein $\sqrt{q}$ depends on the extrinsic time parameter $p$, and $k$ too appears in the potential of the scalar field Hamiltonian. This time-dependence of the Hamiltonian is the consequence of an internal clock which arose from a d.o.f. of the theory.

  While consistent extension of the physical reduced Hamiltonian of GR is a significant aspect of this work, other approaches to resolve the problem of time and extract the physical d.o.f. have been tried before.  Reference \cite{Anderson} starts with a Baierlein-Sharp-Wheeler action\cite{BSW} (in Ref.\cite{SOOYU} this is also discussed within the context of intrinsic time gravity); and the main distinction is that instead of using $\ln q^{\frac{1}{3}}$ as time and solving the Hamiltonian constraint though the elimination of $\tilde\pi$,  an extra $\lambda$  time parameter is introduced with resultant constraint in the form of the Lichnerowicz-York equation. In Ref.\cite{Hanson-Regge-Teitelboim}, transverse-traceless physical decomposition was carried out with Minkowski background for the spatially noncompact case, while York's method was discussed for spatially compact manifolds. Our work demonstrates explicit transverse-traceless decomposition for generic backgrounds and we also compared the reduced Hamiltonians of extrinsic and intrinsic time formulations earlier. Functional path integral for the gravitational field has been developed earlier by Teitelboim by analogy with the quantum mechanics of covariant relativistic point particle\cite{Teitelboim}, whereas our work formulates GR without general covariance. In Refs.\cite{Giesel} and \cite{Husain} additional relativistic dust matter was invoked. Besides current observational conformity with cold, rather than relativistic, dark matter, the relative sign difference with the ${\tilde\pi}^2$ term in (\ref{total}) is, as discussed in scalar field time, a potential problem in guaranteeing the reality of the square-root in the reduced Hamiltonian. This may be curbed by imposing suitable energy conditions; whereas the use of York extrinsic time or our intrinsic time variable is an alternative strategy which exploits the sign difference to overcome the problem.

\section*{Acknowledgments}
This work has been supported in part by the U.S. Naval
Academy, the Ministry of Science and Technology (R. O.
C.) under Grants No. MOST 105-2112-M-006-010 and
No. 106-2112-M-006-009, and the Institute of Physics,
Academia Sinica. E. E. I. would also like to acknowledge
the support of the University of South Africa (UNISA),
Department of Mathematical Sciences, and to express his
gratitude for the hospitality under the Visiting Researcher
Program.

\end{document}